\documentclass[USenglish,twocolumn]{article}

\usepackage[utf8]{inputenc}				
\usepackage[big,online]{dgruyter}	
\usepackage{lmodern} 
\usepackage{microtype}
\usepackage[numbers,square,sort&compress]{natbib}
\usepackage{bm}
\usepackage{glossaries}
\newacronym{CDOS}{CDOS}{cross density of optical states}
\theoremstyle{dgthm}

\usepackage{changes}
\theoremstyle{dgdef}

\newcommand{\bfsfI}{\mbox{\sffamily\bfseries{I}}}

\begin{document}

\articletype{Research Article}
\startpage{1}

\title{Collective single-photon emission and energy transfer in thin-layer dielectric and plasmonic systems}
\runningtitle{Collective single-photon emission and energy transfer in thin-layer systems}

\author[1]{Mads A. J{\o}rgensen}
\author[1]{Devashish Pandey}
\author[2]{Ehsan Amooghorban}
\author[1,3,4]{Sanshui Xiao} 
\author[1,3,4]{Nicolas Stenger} 
\author*[1,3,4]{Martijn Wubs}
\runningauthor{M. A. J{\o}rgensen et al.}
\affil[1]{\protect\raggedright 
Department of Electrical and Photonics Engineering, Technical University of Denmark, Kgs. Lyngby, Denmark, mwubs@dtu.dk (MW)}
\affil[2]{\protect\raggedright 
Department of Physics, Faculty of Science, Shahrekord University, P.O. Box 115, Shahrekord 88186-34141, Iran}
\affil[3]{\protect\raggedright 
Center for Nanostructured Graphene, Technical University of Denmark, Kgs. Lyngby, Denmark}
\affil[4]{\protect\raggedright 
NanoPhoton—Center for Nanophotonics, Technical University of Denmark, Kgs. Lyngby, Denmark}
	
	
\abstract{
We study the collective photon decay of multiple quantum emitters  embedded in a  thin high-index dielectric layer such as hexagonal boron nitride (hBN), with and without a metal substrate.
We first explore the significant role that guided modes including surface plasmon modes play  in the collective decay of identical single-photon emitters (super- and subradiance).
Surprisingly, on distances relevant for collective emission, the guided or surface-plasmon modes do not always enhance the collective emission. We identify configurations with inhibition, and others with enhancement of the dipole interaction due to the guided modes.  We interpret our results in terms of local and cross densities of optical states. 
In the same structure, we show a remarkably favorable configuration for enhanced F\"orster resonance energy transfer between a donor and acceptor in the dielectric layer on a metallic substrate. We compare our results to theoretical limits for energy transfer efficiency.
}


\maketitle

\section{Introduction}

The cooperative emission from multiple emitters is the foundation of promising new technologies like superradiant lasers~\cite{Holland2009a,Bohnet2012,Leymann:2015a,Nefedkin:2017a,Protsenko:2021a}, single-photon emission~\cite{Black:2005a} or quantum memory for information processing~\cite{Trebbia2022}. 
Super- and subradiance are the respectively faster or slower emission of light from multiple emitters due to a shared coupling to the electromagnetic environment. It was originally described in a many-emitter, many-photon limit  in free space~\cite{Dicke1954a,Gross:1982a,Scheibner2007,Svidzinsky2010a,SokolovKupriyanovHavey:2011a,Angerer2018}. 
For multi-emitter technology, it is also important to study another limit of collective emission, namely  a single photon shared between few emitters~\cite{MilonniKnight:1974a,Scully:2009a,Jorgensen:2022a}, in an engineered photonic environment. 

Superradiance in solid state systems is an active field of study~\cite{Sivan2019a,Carminati2019a,Carminati2022a,Reitz2022a,MiguelTorcal:2024a,Pandey:2024a} and often involves waveguide geometries~\cite{Albrecht:2019a,Grim:2019a,Berman:2020a,HuLuLuZhou:2020a,Tiranov:2023a}. %
Here we will consider quantum emitters in Van der Waals materials or subwavelength thin films, which by themselves already are planar waveguides. For distant emitters along one-dimensional waveguides,  waveguide modes dominate the collective emission~\cite{Grim:2019a,Tiranov:2023a}, but for our planar waveguides the situation is less clear. 

We study how collective emission could be enhanced or suppressed by embedding the thin film that contains the quantum light sources  into other planar photonic waveguide    structures. As our main example, we will consider strongly subwavelength-thin films of hexagonal boron nitride (hBN), which  has a large band gap of around 6~eV and can host various types of single-photon emitting color centers ~\cite{Tran:2016a,Mendelson2019a,Vogl2019a,Fischer:2021a,fournier2021position,Aharonovich:2022a,Fischer:2023a}. Some of these have narrow emission lines even at room temperature~\cite{Tran:2016a} and can be deterministically localized~\cite{Chi2021a}. 

Well known challenges for observing collective emission in solid-state environments include inhomogeneous broadening, spectral diffusion, and decoherence due to phonon interactions~\cite{ Cong:2016a,Pandey:2024a}. Nevertheless, collective light emission has been observed in various solid-state environments~\cite{Cong:2016a,Grim:2019a,Biliroglu:2022a,Tiranov:2023a}. 
Decoherence due to phonons can be reduced by working at lower temperatures, although cryogenic temperatures by themselves are no guarantee for lifetime-limited operation of color centers in hBN~\cite{AharonovichWhite2021a}.  
Recently electric-field modulation was shown to actively reduce spectral diffusion and to tune emission lines and narrow  linewidths of color centers in hBN down  to almost the homogeneous lifetime linewith~\cite{Akbari:2022a}. This may also bring closer  the observation of collective emission by color centers in hBN. Other promising emitters in hBN are so-called blue color centers, which can be produced deterministically by electron beams and show surprisingly little inhomogeneous broadening~\cite{fournier2021position}.

Inspired by these developments, we study how collective emission of single photons can be enhanced or suppressed among resonant lifetime-limited color centers in hBN, by engineering their interaction. Besides this idealized coherent limit, where collective emission takes place, we will also consider the opposite incoherent limit, where energy exchange between embedded emitters is a one-way energy transfer process. 
In both limits, we ask how the layered environment, with its associated guided modes, affects the interaction between emitters. 

\begin{figure}[t!]
         \centering
         \includegraphics[width = 8.5cm, height = 2cm]{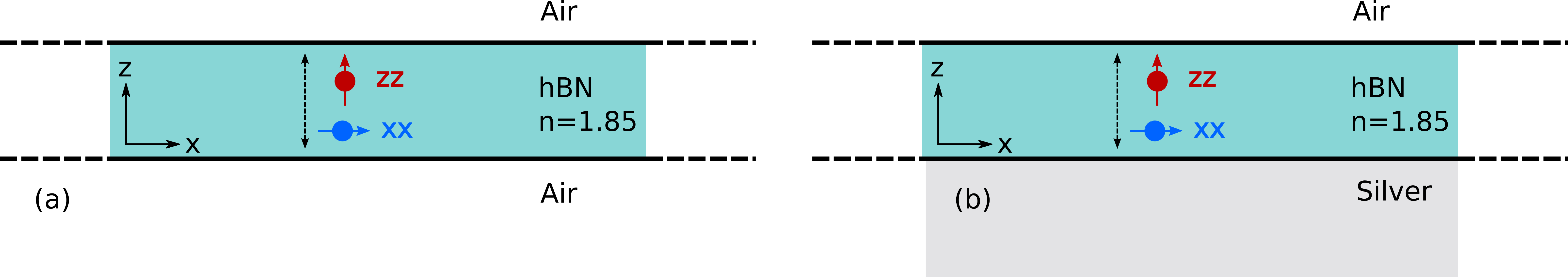}
          \caption{A thin layer of hBN, of thickness $d=\lambda/10$, is surrounded by air on top and by (a) air or (b) silver below. One or more emitters are embedded in the hBN layer. We consider both in-plane and out-of-plane dipole orientations, both corresponding to reported types of emitters in hBN. 
          }
        \label{fig:hBNSetupLDOS}
\end{figure}

Metal interfaces and metallic nano-particles are widely used to enhance light-matter interactions by making use of their  associated surface plasmons~\cite{Chance1974a,GERSTEN1984a,Pustovit2011a,Poddubny2015a,Poudel2016a,Jones:2018a}. Here we report how interactions between embedded quantum emitters are influenced by positioning the thin film onto a metal substrate, as sketched in Fig.~\ref{fig:hBNSetupLDOS}. 

We will also consider the role of  dipole orientations; usually dipoles in 2D materials point in the  plane of the atomic layers, but out of plane dipoles have also been reported~\cite{Dietrich2020a,Hoese2020a}. 

In Sec.~\ref{Sec:Theory} we introduce basics of light emission in terms of the classical Green function, and how to calculate the latter for a multilayer medium. In Sec.~\ref{Sec:Emission_in_layers} we calculate single-emitter decay rates, inter-emitter interactions, and collective emission rates. We isolate the contribution of the guided modes to the total interaction and find both enhanced and suppressed inter-emitter interaction due to the interference of guided modes with the radiative modes. We interpret these results in terms of the concepts of local density of optical states (LDOS) and cross density of optical states (CDOS). Subsequently, we calculate super- and subradiant decay rates 
for emitters in the top of the hBN layer and find both enhanced collective decay at finite emitter separations as well as a longer range for collective emission.  For the same multilayer geometry, we also study energy transfer rates (incoherent limit) that are governed by the same classical Green function, and we find a donor-acceptor configuration with a surprisingly efficient resonant energy transfer. We compare this with theoretical limits for the energy transfer efficiency.
We end with a discussion and conclusions. 

\section{Emission, transfer, and Green function}\label{Sec:Theory}
In this section we introduce the important concepts needed to analyze spontaneous and collective emission, as well as energy transfer in the layered media that we will consider. 

{\em Single-atom decay rates} and local densities of optical states (LDOS) have been calculated for layered systems before~\cite{DREXHAGE1970,Chance1974a,Hoekstra2005a,Palacino2017a}. 
The decay of an emitter depends on its electromagnetic environment and the response to light of any given medium is encoded in the corresponding classical electromagnetic  Green function $\overleftrightarrow{\mathbf{G}}$ (a tensor) of the medium, defined by 
\begin{equation}\label{eq:DyadicGreenFunctionDefiningEquation}
    \left(\epsilon(\mathbf{r},\omega)\frac{\omega^2}{c^2}\bfsfI-\nabla\times\nabla\times \right)\overleftrightarrow{\mathbf{G}}(\mathbf{r},\mathbf{r}',\omega)=\delta(\mathbf{r}-\mathbf{r}')\bfsfI,
\end{equation}
where $\epsilon(\mathbf{r},\omega)$ is the inhomogeneous dielectric function (here assumed scalar) and $\bfsfI$ is the $3\times 3$ identity matrix.
The spontaneous-decay rate $\Gamma$ of an emitter with dipole moment $\bm \mu$ at position ${\bf r}$ radiating at frequency $\omega$ is given by
\begin{equation}\label{eq:SingleEmitterDecayRate}
    \Gamma = 2 \gamma = -\frac{2 \mu^2 \omega^2}{ \hbar \epsilon_0 c^2} \hat{\bm{\mu}} \cdot\text{Im}\left[ \overleftrightarrow{\textbf{G}}(\mathbf{r},\mathbf{r},\omega)\right]\cdot \hat{\bm{\mu}},
\end{equation}
in terms of the imaginary part of the Green function. This is a self-interaction in the sense that the Green function is twice evaluated at the same  spatial point $\mathbf{r}$, in our case a point within the hBN layer of Fig.~\ref{fig:hBNSetupLDOS}. This intensity  decay rate  $\Gamma$ equals twice the field amplitude decay rate $\gamma$ and is proportional to the partial LDOS~\cite{novotny_hecht_2012,Amooghorban:2024a} (here called PLDOS): 
\begin{equation}\label{eq:LDOSasFunctionOfImG}
    \text{PLDOS}= - \frac{6\omega}{\pi c^2}\hat{\bm \mu}\cdot \text{Im}\left[\overleftrightarrow{\textbf{G}}(\textbf{r},\textbf{r},\omega)\right] \cdot\hat{\bm \mu} = 3\sum_\lambda  \left| \textbf{f}_\lambda(\textbf{r})  \cdot \hat{\bm \mu} \right|^2 \delta(\omega - \omega_\lambda).
\end{equation}
It follows immediately that $\Gamma = (\mu^2 \omega \pi/(3\hbar \varepsilon_0 )) {\rm PLDOS}$. This PLDOS would turn into the full LDOS by averaging over dipole orientations, but we will not do that here since we consider  fixed dipole orientations. The last equality in 
Eq.~(\ref{eq:LDOSasFunctionOfImG}) features a sum over a complete set of optical eigenmodes $\textbf{f}_\lambda(\textbf{r}) $ of the dielectric environment, and shows explicitly that the PLDOS is a non-negative quantity. Such a complete eigenmode expansion is only possible for real-valued non-dispersive dielectric functions.   

A number $N$ of quantum emitters does not always emit light at their respective single-atom decay rates, however. Collective emission can take place if the emitters are (nearly) resonant and their interactions strong enough to overcome both detuning and dephasing. The latter requires them to be either in each other's near field and/or they should be connected by a waveguide. The decay dynamics will then be described by $N$ collective decay rates that depend both on the single-emitter decay rates, the detunings between the emitters, and on the pairwise interaction strengths between the emitters. The emitter-emitter interaction is described by the full classical Green function (i.e. both its real and imaginary parts) that connects the locations of the pair of emitters. In an inhomogeneous  nanophotonic environment it is not necessarily so that all emitters have equal transition frequencies and equal single-atom decay rates. But if this is the case, then the collective decay rates that are larger (smaller) than the single-atom decay rate can be called superradiant (subradiant), analogous to the naming for free space.  

Therefore, in order to determine  {\em collective emission rates}, we need to quantify the full interaction, 
of emitters in a complex environment. This is more challenging than calculating the PLDOS, due to the highly singular behaviour of the real part of the Green function in the near field. 
Spatially separated emitters can interact via a shared coupling to the electromagnetic environment. 
Using a multiple-scattering formalism that takes into account all scattering events between any number of emitters~\cite{Wubs2004a,Jorgensen:2022a}, or a master-equation formalism~\cite{Lehmberg:1970a,Chang:2018a,Pandey:2024a},  the collective modes of two quantum emitters in a photonic environment can again be expressed in terms of the Green's function $\overleftrightarrow{\bm G}$ of that environment. 

So that is how we will calculate the interaction between dipoles in the thin films, recall Fig.~\ref{fig:hBNSetupLDOS}. We will make the simplifying assumptions of two identical and lifetime-limited emitters, positioned at $\textbf{r}_1$ and $\textbf{r}_2$ at equal heights. This equal-height assumption is  good for color centers generated by irradiation as in Refs.~\cite{Vogl2019a,Fischer:2021a}. With these assumptions, their single-atom emission rates become identical. (In practice, transition frequencies of two emitters will differ and need external tuning to be brought into resonance.) 
For such a pair of quantum emitters, the complex frequencies of the super- and subradiant modes are found as 
\begin{eqnarray} \label{eq:2AtomSuperradiantResonanceFrequencies3}
    \omega^{\pm}=\tilde{\Omega} -i \gamma
    \pm J_{12}(\tilde{\Omega}),
\end{eqnarray}
where
\begin{equation}\label{eq:interatomic_interaction}
    J_{12}(\omega)=\frac{\mu^2 \omega^2}{ \hbar \epsilon_0 c^2} \hat{\bm{\mu}}_1 \cdot \overleftrightarrow{\textbf{G}}(\mathbf{r}_1,\mathbf{r}_2,\omega)\cdot \hat{\bm{\mu}}_2
\end{equation}
is the complex-valued inter-emitter interaction
and $\tilde{\Omega}$ is the phenomenologically observable emission frequency into which the self-interaction has been absorbed. The generalization of Eq.~(\ref{eq:2AtomSuperradiantResonanceFrequencies3}) to detuned transition frequencies $\tilde{\Omega}_1 \ne \tilde{\Omega}_2$ and/or unequal single-atom emission rates $\gamma_1 \ne \gamma_2$ wil not be considered here but can be found in Refs.~\cite{Wubs2004a,Jorgensen:2022a,Tiranov:2023a}. 

The electromagnetic Green's function appears as the mediator of the electromagnetic field. It encodes all of the information about the response of the environment in which the emitters are treated as scatterers.
From Eq.~(\ref{eq:2AtomSuperradiantResonanceFrequencies3}) it is clear that the real and imaginary parts of $J_{12}$ play fundamentally different roles: the real part constitutes a shift in emission frequency, sometimes called the collective Lamb shift, while the imaginary part of $J_{12}$ influences the two collective decay rates
\begin{equation}\label{eq:2AtomSuperradiantDecayRatesIdentical}
    \gamma^\pm=-\text{Im}[\omega^\pm] = \gamma \mp \text{Im}\left[J_{12}\right].
\end{equation}
Here the interaction $J_{12}$ is not a self-interaction, i.e. the Green function in Eq.~(\ref{eq:interatomic_interaction}) describes the propagation of light between two different emitters. Its imaginary part ${\rm Im}[\overleftrightarrow{\textbf{G}}(\mathbf{r}_1,\mathbf{r}_2,\omega)]$ is proportional to the partial cross density of optical states, or partial CDOS~\cite{Carminati2019a}, here further abbreviated as PCDOS, 
\begin{equation}\label{eq:PCDOS}
    {\rm PCDOS}=-\frac{6 \omega}{ \pi c^2} \hat{\bm{\mu}_1} \cdot\text{Im}\left[ \overleftrightarrow{\textbf{G}}(\mathbf{r}_1,\mathbf{r}_2,\omega)\right]\cdot \hat{\bm{\mu}_2},
\end{equation}
which is a measure of the number of electromagnetic modes connecting two points, per energy~\cite{Carminati2019a,Carminati2022a}, and becomes a PLDOS if the two  positions and dipole moments become equal. Just like the PLDOS, the PCDOS can be expanded into a sum of a complete  set of optical modes~\cite{carminati_schotland_2021},
\begin{equation}\label{eq:CDOSasSumOverModes}
    \text{PCDOS}= 3\sum_\lambda  \hat{\bm \mu}_1 \cdot \text{Re}\left[  \textbf{f}_\lambda(\textbf{r}_1) \textbf{f}^*_\lambda(\textbf{r}_2 ) \right] \cdot \hat{\bm \mu}_2  \delta(\omega - \omega_\lambda).
\end{equation}
So the PCDOS adds up how all the modes connect two unit vectors $\hat{\bm \mu}_{1,2}$ in two points $\textbf{r}_{1,2}$.  
The PCDOS plays an important role in the collective decay of two emitters with identical observable emission frequencies, and satisfies the identity $\text{PCDOS} = -6\hbar \varepsilon_0/(\mu^2 \omega \pi)\text{Im}\,\left[J_{12}\right]$. 

Super- and subradiance can occur when quantum coherences survive on the time scale of spontaneous emission. In the opposite incoherent limit, energy transfer between emitters can still take place, by F{\" o}rster resonance energy transfer (FRET). FRET is the process of 
energy transfer between two particles through the exchange of virtual photons in the near-field of the emitters. Typical inter-emitter separations are a few nanometers~\cite{Foerster1948a,ANDREWS1989195,Andrew:2000a,Lakowicz2006a}. FRET 
goes from an initially excited  donor emitter to a lower-energy acceptor, with broad overlapping emission and absorption spectra. 
 
The classical Green function describes not only superradiance but also the  one-way FRET rate $\Gamma_{\rm FRET}$ from the donor (D) to the acceptor (A), as~\cite{Dung:2002a}
\begin{equation}\label{eq:foerster_theory}
    \Gamma_{\rm FRET} = \int_{-\infty}^{\infty}\mbox{d}\omega\, \sigma_{\rm DA}(\omega) | J_{\rm DA}(\omega,\bf{r}_{\rm A},\bf{r}_{\rm D})|^2.
\end{equation}
Here the spectral overlap function $\sigma_{\rm DA}(\omega)$ is proportional to the product of the donor emission spectum $\sigma_{\rm D}(\omega)$ and the acceptor absorption spectrum $\sigma_{\rm A}(\omega)$.  In general, there is no direct proportionality between the LDOS  and the FRET rate~\cite{Wubs:2015a,Rustomji2021a}.

An excited donor emitter either emits spontaneously or transfers its photon energy to the acceptor. The corresponding FRET efficiency $\eta_\text{FRET}$ is given by
\begin{equation}\label{eq:FRETEfficiency}
  \eta_\text{FRET}=\frac{\Gamma_\text{FRET}}{\Gamma_\text{FRET} + \Gamma^D} = \frac{\Gamma_\text{FRET}}{\Gamma_\text{FRET} + \Gamma^D_\text{rad} + \Gamma^D_\text{nrad}},
\end{equation}
where $\Gamma_\text{FRET}$ is the total FRET transfer rate between the donor and acceptor, and $  \Gamma^D= \Gamma^D_\text{rad}+\Gamma^D_\text{nrad}$ is the total  decay rate of the donor, including the radiative and non-radiative decay rates, respectively. For comparison, the fluorescence quantum yield of the donor is obtained from Eq.~(\ref{eq:FRETEfficiency}) by replacing $\Gamma_\text{FRET}$ in the numerator by $\Gamma^D_\text{rad}$, which shows that FRET efficiency and quantum yield are efficiencies of competing processes that add up to unity if $\Gamma^D_\text{nrad}=0$ (and note that $\Gamma^D_\text{nrad}$ does not contain $\Gamma_\text{FRET}$ in our definition).  

Eq.~(\ref{eq:FRETEfficiency}) for the FRET efficiency is only valid in the limit where the excitation jumping back from the acceptor to the donor can be neglected, that is when $\Gamma^A\gg \Gamma_\text{FRET},\Gamma^D$.
If the transfer is to be efficient, the interaction has to be large enough to compete with single-emitter decay of the donor. 
Both the  FRET rate and the FRET efficiency increase if the spectral overlap of the donor and acceptor emitters grows. Below  we propose instead a specific geometry that presents an alternative way to increase the FRET efficiency in Eq.~(\ref{eq:FRETEfficiency}), namely by suppressing the decay rate of the donor while keeping the total FRET rate  high. 

\section{Results}\label{Sec:Emission_in_layers}

In the following, we will compare two three-layer structures, the first one consisting of a thin layer of hBN ($\epsilon_\text{hBN}=1.85^2=3.4225$ and thickness $d=\lambda/10$~\cite{Rah:2019}) in air ($\epsilon_\text{air}=1$). The second structure consists of the same thin hBN layer now placed in between air and a silver substrate  with $\epsilon_\text{Ag}=-13.529 + 0.416 i$~\cite{JohnsonChristy1972a}. 
These dielectric constants are tabulated values at $\lambda=560\,{\rm nm}$, since  hBN can host bright quantum emitters at this wavelength with a corresponding photon energy of 2.2~eV~\cite{Tran:2016a,Fischer:2021a,Fischer:2023a}.

We will calculate position- and frequency-dependent single-emitter decay rates described by Eq.~(\ref{eq:SingleEmitterDecayRate}), collective decay rates as in Eq.~(\ref{eq:2AtomSuperradiantResonanceFrequencies3}), and energy-transfer rates according to Eq.~(\ref{eq:foerster_theory}). The unknowns in these equations are related to the classical Green function for the layered systems considered. We will calculate these Green functions as integrals over in-plane wavevectors, relegating most technical details to appendices, and separate out the effects of guided and surface-plasmon modes from the other electromagnetic modes.

\subsection{Single-emitter decay in a thin slab} \label{sec:SingleEmitterInSlab}

\begin{figure*}[t!]
         \centering
         \includegraphics[width=\textwidth]{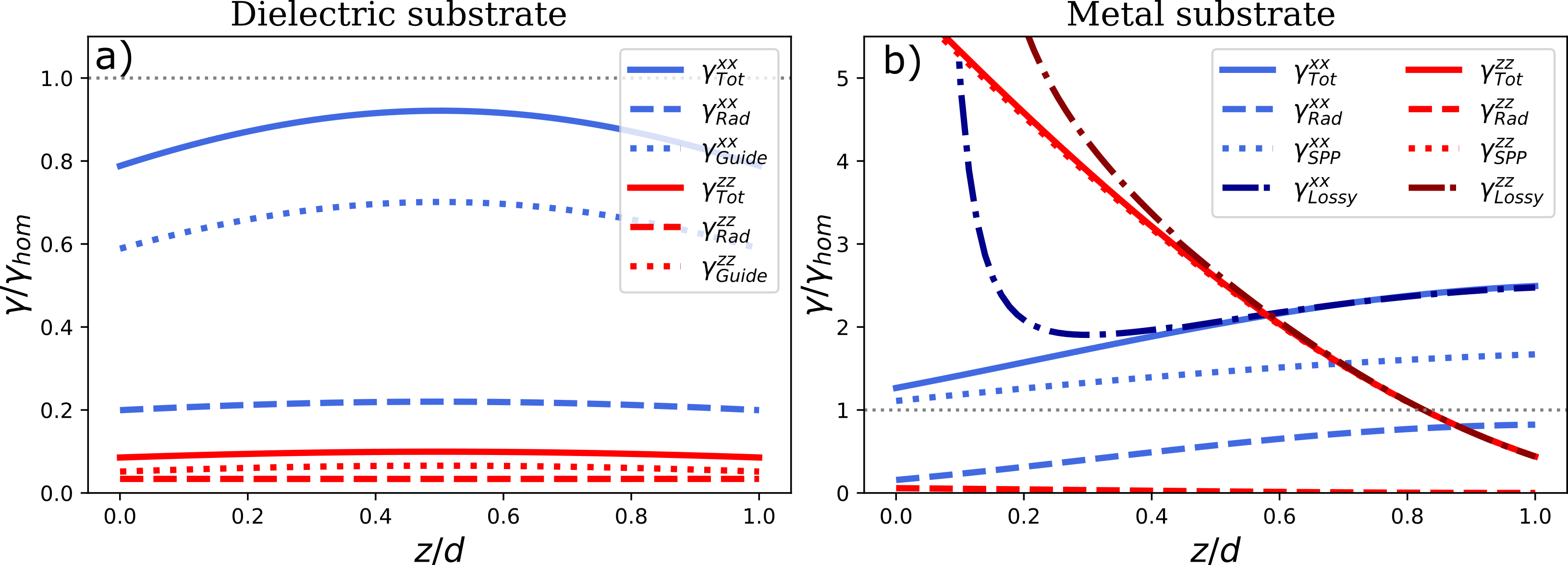}
           \caption{Decay rate of a single emitter at a wavelength $\lambda=560$ nm as a function of scaled height $z/d$ in an hBN layer of thickness $d =\lambda/10$ for (a) an  air/hBN/air and (b) an  air/hBN / silver three-layer system. The value $z=d$ corresponds to the top hBN-air interface. Blue (red) curves correspond to in-plane (out-of-plane) dipoles. The total decay rates are shown in solid lines, while combined contributions of guided and SPP modes are shown in dotted lines. The differences between solid and dotted lines are shown as dashed lines and labeled as ``radiative'' contributions. In (b), 
        the dark blue and dark red dash-dotted lines are total decay rates for silver ($\epsilon=-13.529 + 0.416 i$), while
        the light blue and bright red curves correspond to ''lossless silver'' (with real-valued $\epsilon=-13.529$).  
        }
        \label{fig:LDOSOverHomgeneousDecayRates}
\end{figure*}

Before studying collective emission in multilayers, for comparison we first study single-atom emission in the same geometries. 
So we first consider a lifetime-limited single dipole emitter embedded inside the central hBN layer. Its decay rate will  be orientation-dependent and the three-layer Green's function that is to be inserted into Eq.~(\ref{eq:SingleEmitterDecayRate}) will reflect this. We will only consider fully in-plane or fully out-of-plane dipole orientations. Both of these high-symmetry cases have been observed for color centers in hBN~\cite{Takashima2020a,Dietrich2020a}.
Fig.~\ref{fig:LDOSOverHomgeneousDecayRates}(a) 
shows the spontaneous-decay rate of an emitter inside the hBN layer as a function of the emitter's height within the layer, both for in-plane and for out-of-plane dipole orientations, and normalized by the decay rate in homogeneous hBN. The guided-mode contribution $\gamma_\text{Guide}$ is shown in dotted lines and includes the sum of both the TE and TM modes, which are found at mode indices (defined in App.~\ref{sec:GuidedModesAndDeformedContourIntegral}) $\alpha_\text{Guide}^\text{TE}=1.201$ and $\alpha_\text{Guide}^\text{TM}=1.027$, respectively. An important role in the decay of the emitter is played by the guided modes, with their combined rate $\gamma_\text{Guide}$ typically larger than the rate $\gamma_{\rm rad}$ corresponding to the remaining radiative decay channels. The total single-emitter decay rate $\gamma_\text{Tot}$ is  given by the sum of $\gamma_\text{Guide}$ and $\gamma_\text{Rad}$. For simplicity, we do not consider decay through other non-radiative channels, i.e. not described by the complex dielectric function $\varepsilon$, that may reduce the quantum efficiency of the emitters further. 

Fig.~\ref{fig:LDOSOverHomgeneousDecayRates}(a) shows a strong suppression of emission by the out-of-plane oriented dipole, which emits only about one-tenth as frequently as the same emitter in the homogeneous medium. 
The situation is different when the thin film rests on a metal substrate. 
The decay rate of an emitter in hBN on top of silver, with and without losses, is shown in Fig.~\ref{fig:LDOSOverHomgeneousDecayRates}(b). The presence of the SPP mode ($\alpha=1.69$) enhances the light-matter interaction, resulting in decay rates that are large compared to those of the homogeneous hBN.
The relative contribution of the SPP mode is weakened with increasing distance to the metal interface for the in-plane dipole orientation, but for the out-of-plane dipole, the SPP mode in lossless silver is responsible for nearly all of the decay.
The lossy metal provides a means of single-emitter decay through Ohmic losses for the emitter when it is close. For a Drude metal this energy transfer to the metal occurs at a rate $\propto f(\omega)/d^3$, with a frequency-dependent coefficient $f(\omega)$ that peaks at the surface-plasmon resonance energy of the interface, but also away from this resonance results in  a diverging decay rate as the emitter approaches the metal ''quenching'')~\cite{Chance:1978a}.  However, the Ohmic losses become negligible compared to the other means of decay already for distances beyond a few nanometers between emitter and metal interface.

\subsection{Interactions between two resonant emitters}

 \begin{figure*}[t]
         \centering
         \includegraphics[width=\textwidth]{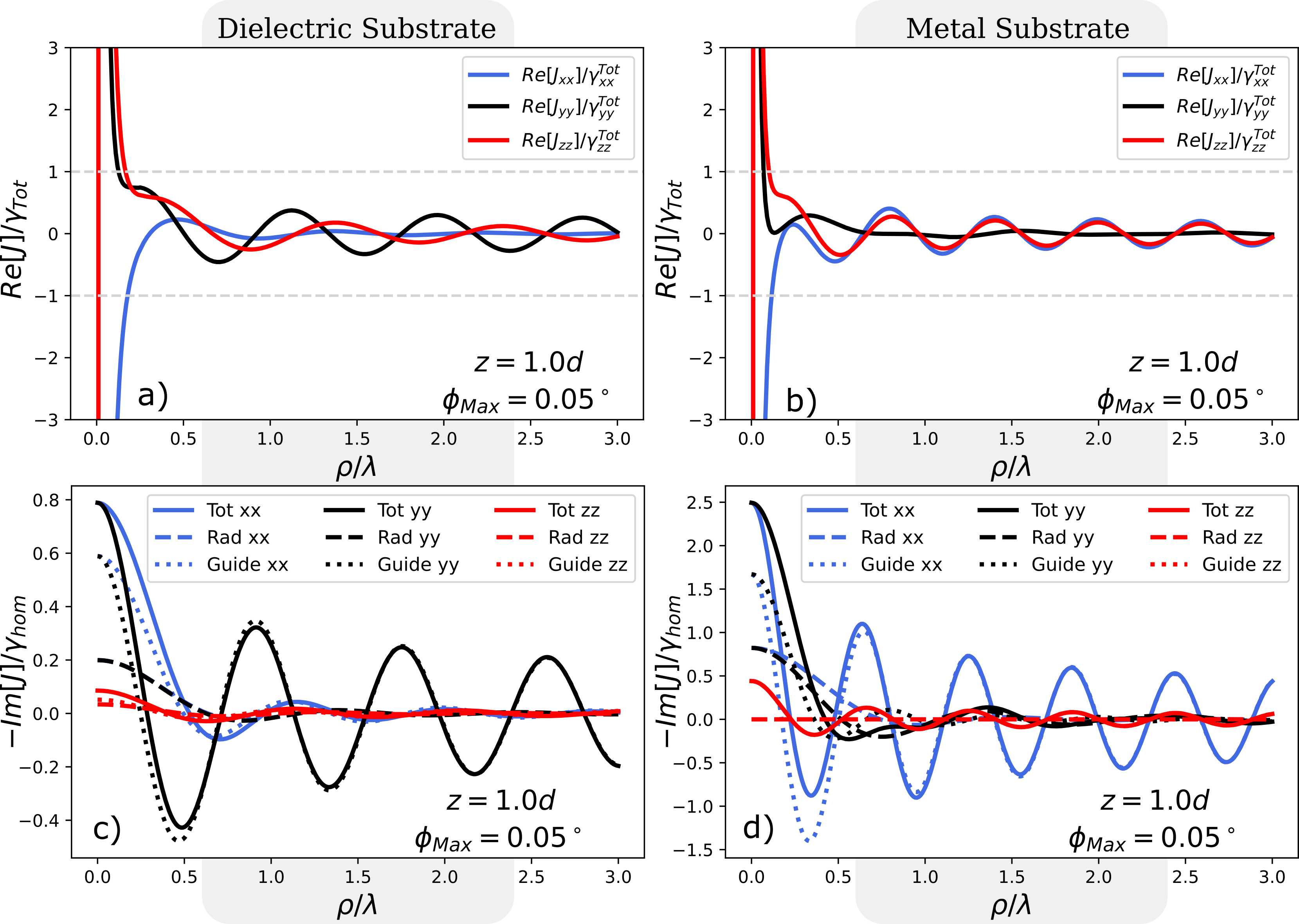}
        \caption{Real and imaginary parts of the  interaction $J_{12}$ between two emitters placed at the top of a central hBN ($\epsilon_{\text{hBN}}=3.4225$) layer of thickness $d=\lambda/10$, as a function of their separation along the $x$-direction. An air substrate is used in panels~(a) and (c) while a (lossless) silver substrate ($\epsilon_{\text{LS}}=-13.529$) is used in  (b) and (d) in order to determine the SPP contribution. In the legend on the right, the total imaginary parts of the interaction are shown in solid while the contributions from the guided modes are shown in dotted lines. The remaining contribution, shown as dashed, is labeled 'rad'. The normalization in (a) and  (b) is the position- and orientation-dependent decay rate of the single emitter in the layered medium, while  (c) and (d) are normalized by the single-emitter decay rate in bulk hBN as shown in Fig.~\ref{fig:LDOSOverHomgeneousDecayRates}. The angle $\phi_{\rm max}$ is related to the numerical calculation as explained in App.~\ref{appendix:Convergence of the contour integration}.}
        \label{fig:hBNInteractionAllNormalizedBygammawg}
\end{figure*}
We consider two emitters in the central layer of hBN, each with dipole moment $\bm \mu_m$ interacting via a single frequency $\omega$ as in Eq.~(\ref{eq:interatomic_interaction}). 
Color centers in hBN produced by ion irradiation tend to exist in the uppermost layers~\cite{Vogl2019a,Fischer:2021a}, and here we will limit ourselves to this case where the emitters are placed at the very top of the hBN layer, $z\approx z_0=d$, furthest from the substrate, whereby we can also model  molecular emitters adsorbed on the surface of the hBN substrate~\cite{Smit:2023a,Neumann:2023a}. 
There have been reports of dipole orientations both in the plane~\cite{Takashima2020a} and out-of-plane~\cite{Dietrich2020a,Hoese2020a}. 
Fig.~\ref{fig:hBNInteractionAllNormalizedBygammawg} shows the real and imaginary parts of the interaction $J_{12}$ between two identical emitters with parallel dipole moments, as a function of the separation along the $\rho=x$-axis. So we define the $x$-direction to point along the line joining the two emitters,  the $y$-direction is the in-plane direction perpendicular to $x$, and $z$ points out of plane.

In order to realize collective emission, the emitter-emitter interaction rate has to become of the order of the single-emitter decay rate. In experiments, the interaction also needs to overcome single-atom dephasing and non-radiative  decay rates, if any, both of which would lead to broadening of the emission spectrum, but we do not consider these additional challenges here.

If one wishes to spectrally resolve the split-peak character of the super- and subradiant modes due to the collective Lamb shift in Eq.~(\ref{eq:2AtomSuperradiantResonanceFrequencies3}), an effect which is characteristic of strongly coupled systems, a Rayleigh criterion of $|\text{Re}[J_{12}]|>\gamma_{\rm tot}$ requires the emitters to be in the near-field, as shown in  Figs.~\ref{fig:hBNInteractionAllNormalizedBygammawg}(a) and \ref{fig:hBNInteractionAllNormalizedBygammawg}(b).
In these two panels (a) and (b) we have made the less common choice of normalizing with the position- and orientation-dependent decay rate, because it allows easy graphical inspection of when the Rayleigh criterion holds. This tells us at what distances collective emission of lifetime-limited emitters is feasible for the configuration studied. 
The equality $|\text{Re}[J_{12}]| =\gamma_{\rm tot}$  is marked by the horizontal grey dashed lines. It can be seen that the Rayleigh criterion is satisfied by emitters approximately at a tenth of the wavelength or less. 

For  $\text{Im}[J_{12}]$ in panels (c) and (d), now scaled by the decay rate of a homogeneous hBN medium, we 
see that the dielectric system in (c) favours collective emission in the $yy-$orientation, in the sense that the position-dependent oscillations in $\text{Im}[J_{12}]/\gamma_{\rm hom}$ have largest amplitudes for those dipole orientations. This can almost fully be ascribed to the coupling between the emitters via the guided mode (dotted curve) of the hBN slab, already for horizontal distances $\rho$ of half a wavelength. The dominance of the waveguide mode at large distances is as expected, while at these short distances it is more surprising. Around $\rho/\lambda = 0.5$ in panel (c), we see that the signs of the guided and radiative parts of $\text{Im}[J_{12}]$ are opposite, resulting in a total $\text{Im}[J_{12}]$ that has a smaller absolute value than its guided part alone. This destructive interference between different channels for interatomic interactions will thus lead to weaker interaction between emitters.

By contrast, the plasmonic system in Fig.~\ref{fig:hBNInteractionAllNormalizedBygammawg}(d) favours collective emission for the in-plane $xx$ dipole orientation, despite the fact that also here considerable `interaction cancellation' happens due to sign differences of the guided and radiative parts of $\text{Im}[J_{12}]$, especially for $\rho/\lambda$ between 0.2 and 0.5.

With the silver substrate present, there is a range of separations where destructive interference between the guided and radiative modes of the in-plane-oriented emitters cause the total interaction to be suppressed. This is most notable for the $yy-$configuration, and means that plasmons are not always beneficial and can be detrimental, not due to their lossy nature, but due to destructive interference. 
This being said,   the interaction between emitters is dominated by the guided modes, both distant and in the near-field. The long-range interaction is mediated almost entirely by the guided modes, illustrated by the merging of the dotted and solid lines in Fig.~\ref{fig:hBNInteractionAllNormalizedBygammawg}.
\begin{figure}[t!]
     \centering
         \centering
         \includegraphics[width=0.45\textwidth]
         {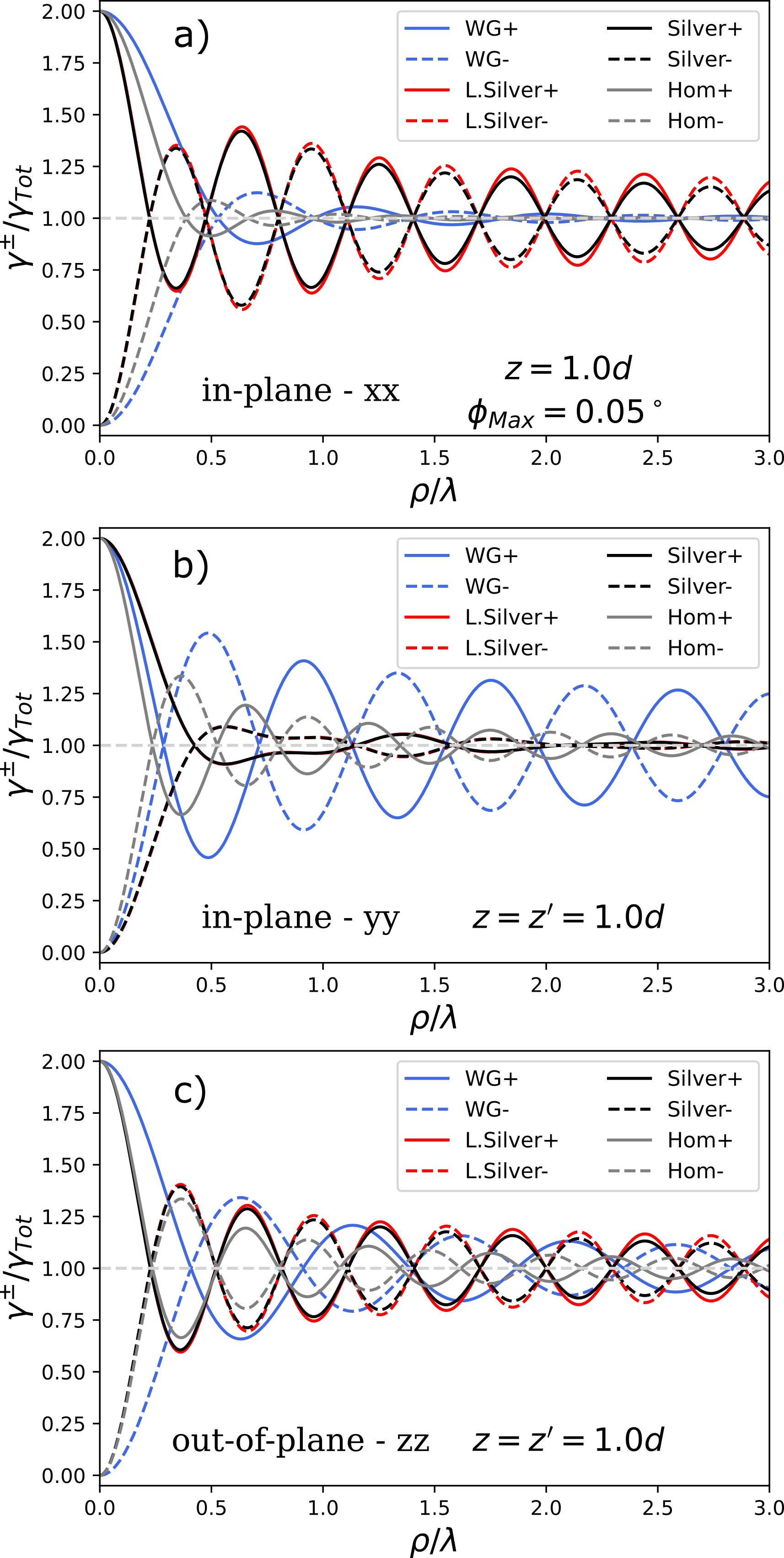}
         \hfill
      \centering
        \caption{Super- and subradiant decay rates of two identical emitters situated at the top of a thin ($d=0.1\lambda$) slab of hBN with air above and either air (blue), lossless silver (red) or lossy silver (black) below. These rates are normalized by the corresponding single-emitter decay rates in the given medium for the same dipole orientation. The axis of separation is along $\bm\hat{x}$ and the dipoles are parallel and oriented along (a) $\bm\hat{x}$, (b) $\bm\hat{y}$ and (c) $\bm\hat{z}$, respectively. The decay rates corresponding to $\gamma^{\pm}=-\text{Im}[\omega^\pm]$ of equation (\ref{eq:2AtomSuperradiantDecayRatesIdentical}) for two emitters 
        are shown in solid (dashed) for plus (minus) modes, respectively. The corresponding collective decay rates of emitters in homogeneous hBN are shown in grey.}
        \label{fig:TwoThreeEmitterSuperradianceIdentical}
\end{figure}

\subsection{Two-emitter superradiance}
The super- and subradiant decay rates of identical emitters are given in Eq.~(\ref{eq:2AtomSuperradiantDecayRatesIdentical}) and shown in Fig.~\ref{fig:TwoThreeEmitterSuperradianceIdentical} for emitters at the top of the hBN layer. The dielectric waveguide (blue), lossy (red) and lossless (black) plasmonic systems are compared to the rates for homogeneous hBN (grey). For all orientations the amplitude of the oscillating collective decay rates in the layered medium surpass those of the homogeneous medium, though the difference is largest for in-plane emitters. 

Comparing the red and black curves in Fig. \ref{fig:TwoThreeEmitterSuperradianceIdentical}, corresponding to the layered air/hBN/silver system with lossless and lossy silver, respectively, we find that Ohmic losses are mostly negligible for collective decay, at least for emitter separations up to a few optical  optical wavelength as shown here. Ohmic losses only really become significant as distances approach the 
propagation length of surface plasmons, which is on the order of multiple microns~\cite{McPeak2015a,Casses:2022a}, or if at least one of the emitters is very close to the metal interface, where its emission may be quenched~\cite{Chance:1978a,CARMINATI2006a,Delga:2014a}, as we mentioned earlier.

Remarkably, for the $xx-$orientation in Fig.~\ref{fig:TwoThreeEmitterSuperradianceIdentical}(a), the   collective rates $\gamma^{\pm}$ show more pronounced peaks and valleys near $\rho/\lambda\approx 0.6$ than for the extrema at the shorter distance $\rho/\lambda\approx 0.35$ (red and black curves). Analogous features occurred in $\rm{Im}[J]$ in Fig.~\ref{fig:hBNInteractionAllNormalizedBygammawg}(d). By contrast, for a bulk medium the analogous collective emission rates for two emitters with equal $z$-coordinates are depicted as the grey curves in Fig.~\ref{fig:TwoThreeEmitterSuperradianceIdentical}(a--c), showing that oscillations in the rates at larger distances are  more damped. Here the reverse situation can occur for the layered plasmonic medium, because of 
 destructive interference between radiative modes on the one hand and guided modes (including surface plasmons) on the other.

For each in-plane orientation of the emitters, there is one type of layered system that exhibits enhanced collective decay for finite emitter separations; the dielectric substrate favours the $yy$-configuration while the silver substrate favours the $xx$-configuration. 
On the other hand, in Fig.~\ref{fig:TwoThreeEmitterSuperradianceIdentical}(b) for separations $\rho/\lambda \gtrsim 0.3$  we see very strong suppression of the collective decay rate of emitters in the $yy$-configuration in the plasmonic system, again due to the cancellation of the radiative and guided mode contributions to the PCDOS. This suppression means that this combination of medium and emitter orientation gives less pronounced collective decay than  emitters in bulk hBN at the same distances, although the latter have no guided modes. 

\subsection{Plasmon-assisted efficient FRET}\label{Sec:efficient_FRET}
Next we examine the one-way F\"{o}rster energy transfer~\cite{Foerster1948a} between two emitters located at different heights within the hBN layer on the silver substrate by utilizing the large differences between the single-emitter decay rates of the donor and the acceptor. 
We are inspired by Ref.~\cite{Pustovit2011a} where it is proposed to  enhance of FRET with the help of localized surface plasmons in metal nanoparticles. 
We will instead consider enhancement of the FRET efficiency helped by the propagating SPPs in our planar geometry.

\begin{figure}[t!]
     \centering
         \centering
         \includegraphics[width=0.45\textwidth]
         {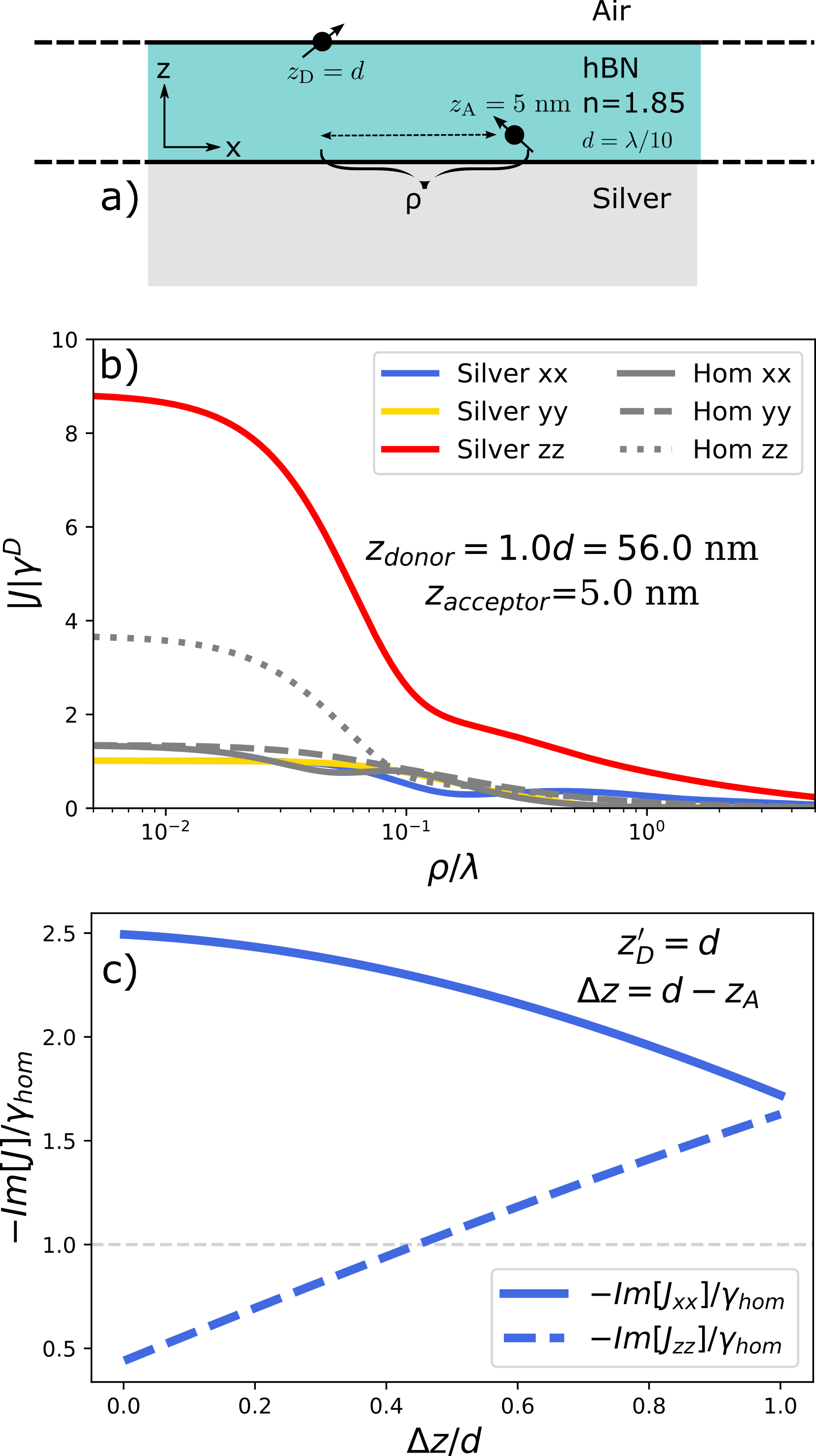}
         \caption{(a) A donor emitter at the upper interface, furthest from the metallic substrate, transfers a photon to the acceptor  located near  the hBN/silver interface. The acceptor is strongly coupled to the SPP mode. (b) Absolute values of the interaction between the donor and acceptor, relative to the donor decay rate, as a function of lateral separation. (c) Imaginary part of the interaction between the donor and the acceptor as a function of acceptor height, with no in-plane separation. The interaction is normalized by the homogeneous hBN decay rate of a single emitter.}
        \label{fig:DonorTopAcceptorBottom}
\end{figure}

The rate of energy transfer goes as the absolute value squared of the Green's function of the medium~\cite{Dung:2002a}, recall Eq.~(\ref{eq:foerster_theory}).  
We choose for the donor to be placed at $z=d$, at the top of the thin film, while the acceptor is placed $z=0.09d=5$ nm above the silver interface in the hBN film, see Fig.~\ref{fig:DonorTopAcceptorBottom}(a). 
These choices could be optimized further, but they will  already show quite efficient energy transfer.  

Referring back to Fig.~\ref{fig:LDOSOverHomgeneousDecayRates}(b), we see that the single-emitter decay rate, or equivalently the PLDOS, is small for out-of-plane emitters that are far from the silver interface, but large for ones that are close.
The donor will therefore indeed have a much smaller spontaneous-decay rate than the acceptor. Interestingly, the interaction between the pair will be many times larger than the donor decay rate for a large range of in-plane separations $\rho$, as illustrated in Fig.~\ref{fig:DonorTopAcceptorBottom}(b).

For in-plane emitters, the interaction is mostly unchanged by the layered medium when compared to homogeneous hBN, but for the out-of-plane dipole configuration, the presence of surface plasmon polaritons has two important effects.  
First, it enhances the donor-acceptor interaction, which leads to a greatly enhanced FRET rate.
Second, it leads to  acceptor decay rates being much larger than the donor decay rates, which makes the energy transfer efficient and one-way.
The combination of both effects makes energy transfer in these plasmonic multilayers viable for  larger  distances than the usual F\"{o}rster range of $\approx10$ nm, see also Refs.~\cite{Baibakov2019a,Hamza:2023a} for alternative schemes.

As seen in Fig.~\ref{fig:DonorTopAcceptorBottom}(b), the FRET rate is highly sensitive to the in-plane separation $\rho$. The maximal FRET rate and efficiency are found for $\rho=0$, i.e. when donor and acceptor share the same $(x,y)$ coordinates. The question is whether this maximal FRET rate also gives a good FRET efficiency. To answer this, we will calculate the efficiency using our Eq.~(\ref{eq:FRETEfficiency}) and compare with the fundamental efficiency bound $\eta_\text{max}=\Gamma_A/(\Gamma_A+\Gamma_D)$ that was derived in Ref.~\cite{Cortes2022a}. Until now we have not specified our donor and acceptor spectra. For a fair comparison, we assume to have the same spectral overlap of two Lorentzian distributions as in Ref.~\cite{Cortes2022a} between the lifetime-limited donor and acceptor emitters, and we made use of Eq.~(7) of Ref.~\cite{Cortes2022a} to compute this.
In doing so, we find a FRET efficiency for the out-of-plane oriented emitters of $\eta_\text{FRET}=0.79$. 
This is quite close to the fundamental efficiency bound, which is $\eta_\text{max}=0.97$ (or $\eta_\text{max}=0.92$ when neglecting loss in silver). 
Our FRET efficiency is also high  compared to some of the configurations considered in Ref.~\cite{Cortes2022a}.

This system is a remarkable case where the imaginary part of the interaction $J_{12}$ between the two emitters is larger than the spontaneous-decay rate of the donor. Or equivalently, the PCDOS is larger than the PLDOS at the donor position. The imaginary part of the interaction is shown in Fig.~\ref{fig:DonorTopAcceptorBottom}(c) as a function of the difference in emitter heights within the hBN layer, $\Delta z=z_D-z_A=d-z_A$. At $\Delta z/d =0$, the PCDOS coincides with the PLDOS at the position of the donor, while the PCDOS between the donor and acceptor emitters for $\rho=0$ in Fig.~\ref{fig:DonorTopAcceptorBottom}(b) corresponds to $\Delta z/d \approx 0.9 $. 
Notably, for the $z$-oriented dipoles, the PCDOS is larger than the PLDOS at the donor position $z_D$ for all vertical donor-acceptor separations in the hBN layer. 

The excited donor emitter will  more likely  transfer its excitation to the acceptor, with the assistance of the SPP mode, than to decay via single-emitter spontaneous decay. Once the energy is transferred, the acceptor will  likely  decay spontaneously, into either heat or into an SPP. The latter could then either be used further in a device, or be detected, for example by converting the SPP into a far-field photon~\cite{Casses:2022a}. 

\section{Discussion: cross density of optical states}\label{Sec:Discussion}

The cross density of optical states (CDOS) was introduced in Ref.~\cite{Caze:2013a} and characterizes  the intrinsic spatial coherence of complex photonic or plasmonic systems. The partial CDOS (PCDOS), defined in Eq.~(\ref{eq:PCDOS}),  
was originally introduced in Ref.~\cite{Carminati2019a}, and controls the (interference term in the) power emitted by two dipoles~\cite{carminati_schotland_2021}. As we have seen, the PCDOS is a useful quantity to understand two-emitter collective emission rates. The same quantity emerges in other contexts, for example in macroscopic quantum electrodynamics as spatial projections of commutators of field operators~\cite{Buhmann:2012a}. The PCDOS has been interpreted as a measure of the number of electromagnetic modes connecting two positions (and orientations), per energy~\cite{Carminati2019a,carminati_schotland_2021,Carminati2022a}. While we do not question the importance of the concept of the (partial) cross density of states, our present study makes us wonder whether the names ``cross density of states'' or ``cross density of optical states'' are appropriate, because of the word `density'.  We will explain this below.

When we  introduced the  PCDOS in Sec.~\ref{Sec:Theory}, we did give the interpretation that it counts the number of eigenmodes connecting two positions (and orientations) for each energy~\cite{carminati_schotland_2021}. And number counting indeed suggests that the concept is related to a density. However, there are important differences between the local  density of optical states (LDOS) and the cross density of optical states.

The PLDOS in Eq.~(\ref{eq:LDOSasFunctionOfImG}) is defined as a sum of non-negative terms, one term for each optical mode. However, there is no corresponding property for the PCDOS in Eq.~(\ref{eq:CDOSasSumOverModes}), since each term in the mode expansion of the PCDOS can be either positive or negative, so cancellations are possible. It is this type of cancellation of the contributions from different modes that we saw in the layered medium, displayed in Figs.~\ref{fig:hBNInteractionAllNormalizedBygammawg}(c,d). So, whereas contributions from individual optical modes are positive and always add up in the PLDOS, the corresponding terms  in the PCDOS can have either sign and may cancel.
As an effect of all modes combined, the PCDOS can switch from  positive to negative and back as we vary the distance between emitters, as we saw in Fig.~\ref{fig:hBNInteractionAllNormalizedBygammawg}. 

Another challenge for the interpretation of the PCDOS as a density or number of modes arose in Fig.~\ref{fig:DonorTopAcceptorBottom} where the PCDOS connecting two points $\textbf{r}_D$ and $\textbf{r}_A$ turned out to be  larger than the  PLDOS evaluated at $\textbf{r}_D$. This complicates the notion that the PCDOS counts a number of modes. For how can there be a larger number of modes connecting two distinct points than the number of modes that pass through just a single one of those points? At least the intuitive interpretation does not hold that  one starts with the PLDOS at point A and then may lose (but not win) a few modes in the PCDOS along the way when moving one of the two positions from point A to B. 

The {\em mode connectivity} is another useful quantity, defined as the PCDOS normalized by the square root of the product of the PLDOS at its two positions~\cite{Carminati2019a}. It does have the interesting property that the mode connectivity of a point A with itself is always unity and larger than the mode connectivity between point A and an arbitrary other point B. However, the mode connectivity is not a density either.     

To summarize this discussion, the PCDOS is an important concept that controls two-atom collective emission and has several other uses. However, the `D' in the acronym stands for `density' and here we gave two arguments why this name may give a wrong impression.

\section{Conclusions}
Knowledge in the nanophotonics community how to engineer single-emitter decay rates is well-developed, while much less is known what nanophotonic  environments are to be preferred to enhance collective emission. We have explored this for plasmonic and dielectric layered geometries.  

We quantified the spectral shifts and collective decay rates of emitters in the experimentally relevant systems of hBN thin films that are freely suspended or placed on a  silver  substrate. In our group we study quantum emitters in hBN flakes~\cite{Fischer:2021a,Fischer:2023a} and  2D materials on top of plasmonic surfaces~\cite{Casses:2024a}. Currently we are not yet able to create nearby quantum emitters in hBN with (almost) identical emission frequencies, but that may change in the future, perhaps enabling collective emission of the type that we studied here theoretically. 

Here we isolated the contributions of the guided modes (including surface plasmons) in single- and multi-emitter decay, and find them to be highly important in mediating the emitter-emitter interaction, even at small emitter separations. The resulting collective decay rates for symmetrically placed emitters in both dielectric and plasmonic systems have sweet spots at finite emitter separation that exhibit a much greater degree of cooperative decay and a longer range than that found in homogeneous media. By `sweet spots' we mean the scaled in-plane distances $\rho/\lambda$ in Fig.~\ref{fig:TwoThreeEmitterSuperradianceIdentical} where the damped oscillatory collective decay rates for planar systems exhibit both maxima on the one curve and minima on the other. The amplitudes of these damped oscillations are much larger than in homogeneous media at similar distances.  For our planar geometries we find collective decay rates that differ by  $40-50\%$ or more from the single-emitter decay rates at emitter separations on the order of hundreds of nanometers. 

We found a strong orientation dependence of both single and collective decay rates. By our definitions, the $x$-direction points along the line joining the two emitters,  the $y$-direction is the in-plane direction perpendicular to $x$, and $z$ points out of plane. For single emitters, rates are strongest modulated for in-plane dipoles in the absence of the metal. With the metal substrate present, perpendicular dipoles point along surface-plasmon field orientations, causing the known strong plasmonic emission-rate  enhancement.  

Surprisingly, as a result of destructive interference between the guided and the radiative decay channels, the first super/subradiant peak at finite distances is weaker than the second one for the plasmon-assisted $xx$-configuration. While plasmonic systems are often used to enhance light-matter interaction, we find that sometimes the SPP mode interferes destructively with the non-guided part of the inter-emitter interaction, on a distance that is important for collective emission in layered geometries. Similar destructive interference does not occur when considering  collective emission by distant identical emitters in 1D waveguides, where the interaction is essentially controlled by the resonant waveguide modes~\cite{Tiranov:2023a}. We do find such destructive interference, which signifies that the cross density of optical states cannot be interpreted as a density.

When combined with life-time limited emitters in hBN, these are promising platforms for realizing collective emission and the technologies that depend upon the collective nature of emission from multi-emitter systems.  
On the other hand, if emission from single emitters is desired, one should take care to suppress  collective emission.

We hope that this work can contribute to the realization of single-photon superradiance in thin films. 
In the future we could take nonlocal response into account~\cite{Gonçalves:2020a}. Furthermore, we treated hBN as an isotropic medium, while it is actually 
 anisotropic due its layered van der Waals nature~\cite{Rah:2019}.
A future study could incorporate this anisotropy in the Green's function using a dyadic permittivity~\cite{Hu:2017a}.

We also examined energy transfer between two asymmetrically placed emitters within hBN, vertically aligned above a metal substrate. This geometry turns out to give very efficient energy transfer, mediated by propagating surface plasmon polaritons. The energy transfer efficiency is close to a theoretical upper bound. After the transfer to the acceptor, the photon either turns into a single surface-plasmon polariton, or into heat. The SPP could be converted into a single photon in a dielectric waveguide, although 1D waveguides may be more ideal for such conversions than our layered geometries. 

More generally, collective emission depends strongly on the dimensionality of the embedding dielectric environment, either in bulk, or in planar or linear waveguides. Here we have investigated  planar structures in more detail, and found destructive interference of guided and other radiative modes in the cross density of optical states that determines collective emission rates.   

\begin{acknowledgement}
We thank Moritz Fischer for stimulating discussions.
\end{acknowledgement}

\begin{funding}
D.P., N.S., and M.W. gratefully acknowledge
support from the Independent
Research Fund Denmark—Natural Sciences (Project No. 0135-00403B). S.X., N.S. and M.W. gratefully acknowledge support from the Danish National Research Foundation through NanoPhoton—Center for Nanophotonics
(Grant Number DNRF147) and Center for Nanostructured Graphene (Grant Number DNRF103).  N.S. and D.P. are supported by the Novo Nordisk Foundation NERD Programme (project QuDec NNF23OC0082957). S.X. acknowledges the support from the Independent Research Fund Denmark (2032-00351B).

\end{funding}

\begin{conflictofinterest}
Authors state no conflict of interest.
\end{conflictofinterest}


\appendix

\counterwithin{figure}{section} 

\section{Multilayer Green function and three-layer system}
The response of a medium is quantified by the dielectric function, $\epsilon(\mathbf{r})$. For a layered medium, the dielectric function is piecewise constant, changing only as a function of $z$ at the interfaces between different layers, recall Fig.~\ref{fig:hBNSetupLDOS}.

Each layer is characterized only by its dielectric constant and its thickness, $d_l$.
Below we briefly recap the properties of the Green function of such layered systems.
The Green's function for a layered medium can be found in various ways. We will make use of the form and notation presented by Toma\v{s} in Ref.~\cite{Tomas:1994a}. This approach uses the homogeneous Green's function in the Weyl representation to find the source field of the emitter embedded in layer $j$ to then find the scattered field in every layer, from which the fully retarded Green's function of the layered medium is extracted. While usually only single-emitter decay rates are calculated with this formalism, we will determine collective emission rates. Alternative approaches can be found in Refs.~\cite{Hartman2001a,Dung2002a,Dung2002b,Hoekstra2005a,Hohenester:2020a}. Using an integral identity of the angular integration, the Green's function for a three-layer medium is given by~\cite{Hohenester:2020a}
\begin{equation}\label{eq:FourierGreens2}
\overleftrightarrow{\textbf{G}}_{lj}(\textbf{r},\omega)=
     \frac{1}{2\pi}\int_0^{\infty}\text{d}k\mkern3mu k   \left<\overleftrightarrow{\textbf{G}}_{lj}(k,\omega,lz,jz_0;\rho,\theta)\right>,
\end{equation}
where $\textbf{r}= \rho \hat{\bm\rho} + z\hat{\textbf{z}}$, $k$ is the magnitude of the in-plane component of the wavevector and $l$ and $j$ represent the layers of the final and initial positions, respectively. The notation $j z_0$ means that the $z_0$-coordinate is taken with respect to the bottom of the $j$'th layer, and similarly for $lz$. For the bottom layer, the $z$-coordinate would be negative. 

For our purposes of color centers within the same thin layer, we only need to consider the Green's function connecting two points within the central layer of a three-layer medium. With $\mathbf{r}$ and $\mathbf{r}'$ both located in layer $l=j=2$, and assuming that $z>z_0$ (where a finite $z$-difference is required for convergence, see Appendix~\ref{appendix:Convergence of the contour integration}), the angle-averaged Green's function in the Weyl representation is given by
\begin{equation}\label{eq:ExampleGreensFunctionIntegrand22z>z0}
\begin{aligned}
 & \left<\right.\overleftrightarrow{\textbf{G}}_{22}(k,\omega,z,z_0;\rho,\theta)\left.\right>  =  \\
 & - \frac{i}{2 \beta_2}\sum_{q=\text{TE,TM}}\frac{\xi_q}{D^q_{2}}\Big[
 \left<\hat{\textbf{e}}^{+}_{q2}(\textbf{k})\hat{\textbf{e}}^{-}_{q2}(-\textbf{k})\right> e^{i \beta_2(z-z_0)} \\
 & +  r^q_{23} \left<\hat{\textbf{e}}^{-}_{q2}(\textbf{k})\hat{\textbf{e}}^{-}_{q2}(-\textbf{k})\right> e^{i \beta_2(-z-z_0 + 2d)} \\
 & +  r^q_{21}\left<\hat{\textbf{e}}^{+}_{q2}(\textbf{k})\hat{\textbf{e}}^{+}_{q2}(-\textbf{k})\right> e^{i \beta_2(z+z_0)} \\
 & +  r^q_{23} r^q_{21} \left<\hat{\textbf{e}}^{-}_{q2}(\textbf{k})\hat{\textbf{e}}^{+}_{q2}(-\textbf{k})\right> e^{i \beta_2(-z+z_0 + 2d)} \Big].
\end{aligned}
\end{equation}

The Green's function for $z<z_0$ can be found through interchange of $\mathbf{r}$ and $\mathbf{r}'$.
In Eq.~(\ref{eq:ExampleGreensFunctionIntegrand22z>z0}), the $\xi_{\text{TM}}=1$ and $\xi_\text{TE}=-1$ are polarization-dependent sign factors, 
   $\beta_{l} \equiv \sqrt{k_l^2-k^2}$
is the pseudo-$z$-component of the wavevector, and the $r^q_{ij}$ are the Fresnel reflection coefficients for $q$-polarization between layers $i$ and $j$.  For the angle-averaged dyadic outer products of the polarization vectors in Eq.~(\ref{eq:ExampleGreensFunctionIntegrand22z>z0}), such as $\left< \hat{\textbf{e}}^{\pm}_{\text{TE}l}(\textbf{k})\hat{\textbf{e}}^{\pm}_{\text{TE}j}(-\textbf{k})\right>$, we refer to Eq.~ (B.8) in Appendix~B of Ref.~\cite{Hohenester:2020a}.

\subsection{Guided modes and the deformed contour integral}\label{sec:GuidedModesAndDeformedContourIntegral}
Of special note in Eq.~({\ref{eq:ExampleGreensFunctionIntegrand22z>z0}) for the multilayer Green function is the multiple-scattering parameter
\begin{equation}\label{eq:Dmultiplescatteringparameter}
    D^q_{l}=1-r^q_{l-}r^q_{l+}e^{2 i \beta_l d_l},
\end{equation}
where $r^q_{l+}$ and $r^q_{l+}$ are the total reflection coefficients of the stacks above and below layer $l$. 
In the form $1/ D^q_{l}$, it accounts for the infinite (geometric) series of reflection events of all orders in the central layer and the phase gained from traversing the $l$'th layer twice. This can be seen by carrying out the Taylor expansion $\frac{1}{1-x}=\sum_{n=0}^\infty x^n$. 
This is true for purely dielectric systems, where the Taylor expansion of $1/ D^q_{l}$ affords a very pictorial description of this multiple-scattering mechanism, but also for metallic systems that can carry SPP modes. 

In general, the zeroes of $D^q_l$ correspond to guided modes in the $l$'th layer. If we let the thickness of the hBN layer tend to zero, then we find a zero of the multiple-scattering parameter for TM-polarized light  that corresponds to the textbook surface plasmon polariton dispersion $k_{\text{SPP}}=k_0 \left[\epsilon_d \epsilon_m/(\epsilon_d + \epsilon_m)\right]^{\frac{1}{2}}$ for a single dielectric-metal-interface~\cite{Maier:2007}, as it should.
No analogous surface-plasmon condition is found for the TE modes, as is well-known for single-interface surface plasmons. Nevertheless, increasing the thickness of the hBN layer will eventually allow for multiple guided modes to form, some of which can be TE polarized. These larger thicknesses are outside the scope of the present paper.\\

In general the mode index, i.e. the value of $\alpha=k/k_0$ for which we have guided modes, has to be found through numerical solution of the condition $D^q_l=0$, at the relevant wavelength of light of interest (here $\lambda = 560$ nm.). For lossless systems, this results in real-valued $\alpha$, while with loss, the poles migrate into the upper half of the complex $\alpha$-plane, as these modes correspond to exponentially damped waves~\cite{Hohenester:2020a}. 
For lossless media it is possible to isolate the contribution of the guided modes that appear on the real axis by evaluating the integral of a small open contour tightly surrounding the pole. Therefore, we will sometimes consider a ``lossless metal'' to isolate the contributions of the guided modes with metallic substrates. This is a useful an accurate approximation for our purposes since finite propagation lengths of SPPs are usually much longer than typical distances between emitters that send out light collectively. For more details on this approximation we refer to Appendix~\ref{sec:ContourIntegrationMethod}.  Ref.~\cite{UrbachRikken1998a} provides a means of counting guided modes for three-layer lossless dielectric systems. In fact, it was shown that for a symmetric three-layer system with lossless dielectric materials, there will always be at least one TE- and one TM-polarized guided mode.

For guided modes in a lossless system,
the residue from the contour integral around the pole is purely real-valued, making the guided-mode contribution to the Green's function purely imaginary via the residue theorem.
From Eqs.~(\ref{eq:interatomic_interaction}),  (\ref{eq:2AtomSuperradiantDecayRatesIdentical}), and ~(\ref{eq:SingleEmitterDecayRate}),  we see that this means that the contribution of the guided modes becomes evident only in the decay rates of single or multiple-emitter systems.
In this work we do not consider each guided mode separately, but instead include all the guided-mode contributions together.

\section{Contour integration}\label{sec:ContourIntegrationMethod}
\begin{figure}[t]
     \centering
      \includegraphics[width = 0.5\textwidth,keepaspectratio]{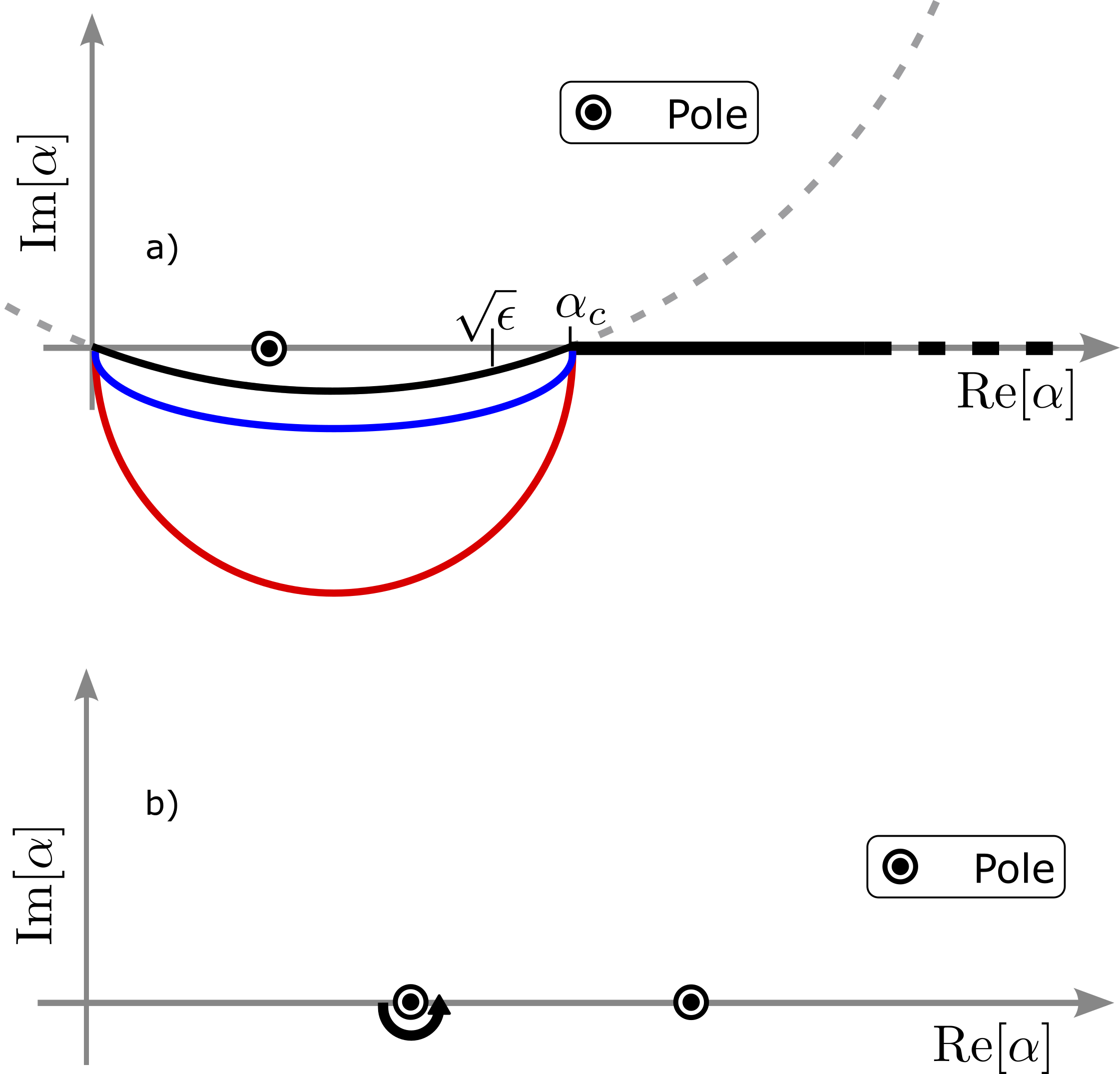}
        \caption{Integration path used to deal with the singularities of poles and the branch points corresponding to $\alpha=\sqrt{\epsilon_l}$. (a) Shown are three different deformed contours that extend into the negative imaginary $\alpha$-plane. The red contour is a semi-circle, the blue is a semi-ellipse and the black is a circle-arc, which the grey dashes showing part of the outline of the bigger circle. After the deformed contour circumvents all problematic points, the integration continues along the real $\alpha$-axis as usual.
        (b) Contribution from a single guided mode via small contour integral.}
        \label{fig:Integrationpaths}
\end{figure}
In order to take into account the contributions of guided modes, which are represented by poles of the Green's function integrand, we partition the integration in Eq.~(\ref{eq:FourierGreens2}) into two parts. In the first, we utilize a deformed integration contour, that goes from $0$ to a cutoff value $\alpha_{\text{c}}$, and then dips into the negative complex plane. This cutoff value has to be large enough that all poles on the real axis have been circumvented. For examples and details, see Refs.~\cite{Chew:1995:Book,Paulus2000a,Hoekstra2005a,Sondergaard2019a,Hohenester:2020a}. The second integral is then performed from $\alpha_{\text{c}}$ to infinity along the real axis.
While a semi-circle path is easy to implement, going far into the fourth quadrant causes exponential decay. A semi-elliptical contour is often recommended in order to better control the depth into the imaginary plane~\cite{Paulus2000a,Hohenester:2020a}. To avoid the large curvatures near the edges of the integration path when the eccentricity of the ellipse is large, we instead make use of a circle arc, shown in black in Fig.~\ref{fig:Integrationpaths}(a).
In order to isolate the contributions of the guided modes in lossless media, we make a tight semi-circle integration around the pole, as illustrated in Fig.~\ref{fig:Integrationpaths}(b). 
Since the pole is located on the real axis, we only pick up the contributions from going half-way around the pole, i.e. only half the residue contributes to the integral. For lossy metals the pole moves into the first quadrant, where branch cuts significantly complicate the process of isolating the contribution from the guided mode~\cite{Chew:1995:Book,Hohenester:2020a}.
Thick layers can be host to multiple guided modes. If these have very similar mode indices, then it can be difficult to isolate their contributions. The same is true for thin layers, where the mode index of the guided mode tends towards the refractive index of the surrounding medium.

\section{Convergence of the contour integration}\label{appendix:Convergence of the contour integration}
In order to get the integral in Eq.~(\ref{eq:FourierGreens2}) to converge, we require a finite difference in the $z$-coordinates of the emitters positions, i.e $|z-z_0|>0$. This can further be justified by the implicit choice of an exclusion volume, as rigorously studied by Yaghjian~\cite{Yaghjian1980a}, for our layered medium. However, we are free to choose a finite, but physically insignificant, value. We chose 
\begin{equation}
    \Delta z= \text{max}\left(\Delta z_\text{min},\rho \sin(\phi_\text{max})\right),
\end{equation}
such that for small separations along the in-plane direction, the vertical separation is constant and bounded from below by $\Delta z_\text{min}$, and for larger in-plane separations the vertical off-set grows with increasing distance.
Our choice of lower bound value is $\Delta z_\text{min}\approx 0.5$~Å, much smaller than other physically relevant length scales. The upper bound is determined by the maximal angle between the actual separation of the points $\textbf{r}$ and $\textbf{r}'$ and the projection onto the horizontal plane, for which we used $\phi_\text{max}=0.05^\circ$, leading to a maximal vertical displacement $|z-z_0|<1.5$~nm for a maximal separation $\rho=3\lambda=1680$~nm.

For the largest distances between the emitters that we consider (three wavelengths), choosing the smallest value of the exclusion volume of 0.5~Å would result in extremely slow convergence of the integral Eq.~(\ref{eq:FourierGreens2}). For smaller distances, we can take the exclusion length smaller without the very slow convergence. With our current choices of exclusion volumes, we do not have this problem of slow convergence, and we checked convergence: our results do not change by making exclusion volumes slightly larger or smaller. 


\end{document}